\documentclass{ws-ijmpcs}

\begin{document}

\markboth{F. P. Poulis \& J. M. Salim}
{Weyl Geometry as Characterization of Space-Time}

%
\catchline{}{}{}{}{}
%

\title{WEYL GEOMETRY AS CHARACTERIZATION OF SPACE-TIME}

\author{F. P. POULIS and J. M. SALIM}

\address{Centro Brasileiro de Pesquisas F\'{\i}sicas, Rua Xavier Sigaud, 150\\
Rio de Janeiro, RJ, CEP 22290-180, Brazil\\
fppoulis@cbpf.br, jsalim@cbpf.br}

\maketitle

\begin{history}
\received{14 June 2011}
\revised{Day Month Year}
\end{history}

\begin{abstract}
Motivated by an axiomatic approach to characterize space-time it is investigated a reformulation of Einstein's gravity where the pseudo-riemannian geometry is substituted by a Weyl one. It is presented the main properties of the Weyl geometry and it is shown that it gives extra contributions to the trajectories of test particles, serving as one more motivation to study general relativity in Weyl geometry. It is introduced its variational formalism and it is established the coupling with other physical fields in such a way that the theory acquires a gauge symmetry for the geometrical fields. It is shown that this symmetry is still present for the red-shift and, considering cosmological models, it opens the possibility that observations can be fully described by the new geometrical scalar field. It is concluded then that this reformulation, although representing a theoretical advance, still needs a complete description of their objects.

\end{abstract}

\ccode{PACS numbers: 04.20.-q, 04.20.Cv, 04.90.+e}

\section{Introduction}
General Relativity concerns a geometrical formulation of gravity and has as a
tenet a pseudo-riemannian geometry for the manifolds taken in consideration.
However, this is an arbitrary choice and is intended to preserve scalar products
 between vectors and, consequently, their lengths when parallel transported along
 an arbitrary path on the manifold.

Nevertheless, considering that in this theory the characterization of space-time
is made by rulers and clocks based on light rays emission and reception, Ehlers,
Pirani and Schild\cite{EPS} (EPS), followed by Woodhouse,\cite{Woodhouse} have shown
that the pseudo-riemannian geometry represents a restriction among the possibilities
those measuring tools allow. That is, starting from axioms concerning light
rays\footnote{Described by a conformal invariant theory, like Maxwell's one.} and
trajectories of freely falling particles you get, necessarily, a Weyl geometry as the
most general one. This, in turn, differs from the pseudo-riemannian one by allowing
vectors to change their modulus when paralel transported along any curve.

Weyl's motivation was to geometrize electromagnetism, with the intend to unify it
with gravitation, and the length variation of vectors was associated with the
electromagnetic potential.\cite{Weyl} Being so, it suffered some criticisms that made
it be abandoned. However, the axiomatic characterization of space-time by EPS and
Woodhouse bring Weyl geometry to light once again. This time there is no association
of length variation with any known physical field. The vector field that performs
such a variation, which we will call $\omega_\mu$, is purely geometrical, and thus
the objections raised before are transcended.

In Sec.~\ref{sec - propriedades} are presented the main properties of Weyl geometry.
From them, preserving the postulate that test particles travel through time-like geodesics,
it is verified that those should be different from the pseudo-riemannian case of General
 Relativity (GR). The $\omega_\mu$ field appears giving extra contributions to the
 trajectories of material particles and thus influencing on an observable of the theory.
 This is one of the motivations to study Einstein gravity in Weyl geometry: as a way to
 determine how or to what extent one can use the new geometric field to make the theory
 more consistent.

In the following section it is made a brief study of congruences of curves in Weyl.
It is defined the concept of physical distance between neighboring curves and it is
shown that their evolution is given by a linear transformation. Next, the matrix of
this transformation is decomposed in their irreducible parts, which are the corresponding
 expressions for the expansion, shear and rotation from GR.

In Sec.~\ref{sec - form varia e acopl} we restrict the Weyl geometry to its integrable
version, where $\omega_\mu \equiv\partial\omega/\partial x^\mu$. In this case, variations
of length depend only on the starting and ending points of the trajectory. Variational
formulations are presented for both obtaining gravitation in this geometry and for
particles to follow geodesics. It is established how other physical fields are coupled
with the new geometric one and it is observed that the coupling used endows our theory
with a gauge symmetry inducing a conformal invariance.

As discussed in Ref.~\refcite{Bekenstein-Conformal invariance}, since Dirac\cite{Dirac}
raised the question about the gravitational constant be, in fact, a constant, this problem
has been a challenge for theoretical physics. However, it is argued that this can be
solved by taking into account the principle that all fundamental equations of physics
must be invariant under local transformations of scale, as proposed by Weyl\cite{Weyl}
and Hoyle and Narlikar.\cite{Hoyle e Narlikar} It is therefore of great theoretical
interest to examine a conformal invariant version of gravitation.

Other versions of a non-conformal model for the gravitational interaction in Weyl
Integrable Space-times (WIST) were already developed\cite{Novello1,Salim-Sautu1} and
applications were made in cosmology\cite{Salim-Sautu2,Salim-Sautu3} and in spherically
symmetric configurations for modeling stars.\cite{Salim-Sautu4,Salim-Sautu5}
Besides that, the WIST geometry has been used to describe the quantum behavior of a
particle in an external potential in the non-relativistic\cite{NST1} and the
relativistic\cite{NST2} regime. Nevertheless, none of them has formulated the theory
in this gauge invariant way.

Carrying on with the reformulation of the theory, it remained to redefine the
observables of GR within this formalism. In Sec.~\ref{sec - red-shift} this is done
for the red-shift and in Sec.~\ref{sec - Verif inv conf} it is shown that it
preserves the gauge symmetry of the theory, as expected. Moreover, in
Sec.~\ref{sec - inv conf em mod cosmo} it is shown that for cosmological models the
conformal invariance opens the possibility that observations may no longer indicate a
curvature nor an evolution to the metric. All observations and conclusions from the
red-shift in GR being fully described in terms of the $\omega$ field.

We conclude that this reformulation of Einstein's gravitation in the more general
and axiomatic context of Weyl geometries turns out to acquire a gauge symmetry that,
although it represents a theoretical advance, introduces an indeterminacy of their
objects which needs to be solved, for consistency.

\section{Some Properties of Weyl Geometry \label{sec - propriedades}}

Weyl's geometry differs from that of Riemann by allowing vectors to also change
their modules when parallel transported along a closed path. This can be summarized
 by the following expression for the covariant derivative of the metric:\cite{Adler}

\begin{equation}
\nabla_\alpha g_{\beta\gamma} = \omega_\alpha g_{\beta\gamma} \;
\Leftrightarrow \; \nabla_\alpha g^{\beta\gamma} = -\omega_\alpha g^{\beta\gamma} \, .
\label{deriv da metrica em Weyl}
\end{equation}

\noindent This implies a connection given by

\begin{equation}
\Gamma^{\alpha}_{\mu\nu} = \hat{\Gamma}^{\alpha}_{\mu\nu} -
\frac{1}{2}\left(\omega_{\mu}\delta_{\nu}^{\alpha} +
\omega_{\nu}\delta_{\mu}^{\alpha} - g_{\mu\nu}\omega^{\alpha}\right) \, ,
\label{conexao de Weyl}
\end{equation}

\noindent where symbols with hat refer to their equivalent in the Riemannian case.

From it, we can also conclude that for any vector $X^\alpha$ we have:

\begin{eqnarray}
& & X_\alpha\nabla_\beta X^\alpha =
\frac{1}{2}\partial_\beta\left( X_\alpha X^\alpha \right) -
\frac{1}{2}X^\alpha X_\alpha\omega_\beta \, ; \label{contracao de X com deriv de X I} \\
& & X^\alpha\nabla_\beta X_ \alpha =
\frac{1}{2}\partial_\beta\left( X_\alpha X^\alpha \right) +
 \frac{1}{2}X^\alpha X_\alpha\omega_\beta \, .
\label{contracao de X com deriv de X II}
\end{eqnarray}

\noindent Then, the most general geodesic equation is given by:

\begin{equation}
u^\beta\nabla_\beta u^\alpha =
\left( \frac{d}{ds}\ln\sqrt{u^\beta u_\beta} \right)u^\alpha -
\frac{1}{2}u^\beta \omega_\beta u^\alpha \, .
\label{Eq geodesica em Weyl}
\end{equation}

\noindent Where $u^\alpha$ is the tangent vector of the geodesic and $s$ is the parameter
 which des\-cri\-bes it.

\section{Congruence of Curves \label{sec - congruencia de curvas}}

We consider a congruence of curves $\gamma(\tilde{s},t)$, $C^2$, which in a local
coordinate system $\left \{x^{\alpha}\right \}$, have coordinates
$x^{\alpha}(\tilde{s},t)$ and we take the parameter $\tilde{s}$ so that

\begin{equation}
d\tilde{s}^2 = e^{-\omega}ds^2 = e^{-\omega}g_{\mu\nu}dx^\mu dx^\nu \, , \quad u^{\alpha}=
\frac{dx^{\alpha}}{d\tilde{s}} \;\; \Rightarrow \;\; g_{\alpha\beta}u^{\alpha}u^{\beta} =
e^\omega \, .
\label{parametro afim}
\end{equation}

Two neighboring points, $P(\tilde{s}_0, t_0) \in \gamma(\tilde{s},t_0)$ and
$Q(\tilde{s}_0, t_0 + \Delta t) \in \gamma(\tilde{s},t_0 + \Delta t)$, define a new
vector which we will call \textit{connection vector}. Their components are
$\left.Z^{\alpha} = \partial x^{\alpha}/\partial t \right|_{\begin{subarray}{l} \tilde{s} =
 \tilde{s}_0 \cr t = t_0\end{subarray}}\Delta t$ and it is associated with the distance,
 on the manifold, between the points $P$ and $Q$.

The physical distance, locally determined by an observer at rest in relation to what
has the curve $\gamma(\tilde{s},t_0)$ as world line, is given by
$\bot Z^\alpha \equiv h^\alpha_\beta Z^\beta$, which we will call \textit{relative
position vector}, where
$h^\alpha_\beta = \delta^\alpha_\beta - e^{-\omega}u^\alpha u_\beta$ makes the projection
of any vector in the space perpendicular to $u^\alpha$.

Introducing the notation
$\dot{X}^\alpha \equiv u^\beta\nabla_\beta X^\alpha \equiv \frac{D}{D\tilde{s}}X^\alpha$
to any vector $X^\alpha$, the relative velocity between two neighboring particles in
the congruence is given by $\bot\frac{D}{D\tilde{s}}\bot Z^\alpha$ and is related to
 $Z^\alpha$ by the following equation:

\begin{equation}
\bot\frac{D}{D\tilde{s}}\bot Z^\alpha =
 h^\alpha_\beta u^\gamma\nabla_\gamma\left ( h^\beta_\delta Z^\delta \right ) =
  h^\alpha_\beta u^\gamma Z^\delta\nabla_\gamma h^\beta_\delta + h^\alpha_\delta\dot{Z}^\delta \, .
\label{velo relativa I}
\end{equation}

By the property $\dot{Z}^\delta = u^\alpha\nabla_\alpha Z^\delta =
 Z^\alpha\nabla_\alpha u^\delta$, we can simplify the result to

\begin{align}
\bot\frac{D}{D\tilde{s}}\bot Z^\alpha & =
 h^\alpha_\beta h^\lambda_\delta \left ( \nabla_\lambda u^\beta  \right ) \bot Z^\delta
 \nonumber \\
& = {V^\alpha}_\beta \bot Z^\beta \, ,
\label{velo relativa III}
\end{align}

\noindent where we have defined
${V^\alpha}_\beta \equiv h^\alpha_\gamma h^\delta_\beta\nabla_\delta u^\gamma$.
This shows that the velocity of separation between neighboring particles is related to
the relative position vector by a linear transformation.

It can be easily verified that
$V_{\alpha\beta} = h^\lambda_\alpha h^\delta_\beta \nabla_\delta u_\lambda$,
then we have their irreducible parts:

\begin{equation}
\left.\begin{array}{rcl}
\theta_{\alpha\beta} & = & V_{(\alpha\beta)}\; ; \quad \theta = {\theta^\alpha}_\alpha \\
\sigma_{\alpha\beta} & = & \theta_{\alpha\beta} - \frac{1}{3}\theta h_{\alpha\beta} \\
\omega_{\alpha\beta} & = & V_{[\alpha\beta]}
\end{array}\right\}\;\Rightarrow\; V_{\alpha\beta} = \omega_{\alpha\beta} +
\sigma_{\alpha\beta} + \frac{1}{3}\theta h_{\alpha\beta} \, .
\label{decomp irredutivel para V}
\end{equation}

\noindent Where $V_{(\alpha\beta)}$ and $V_{[\alpha\beta]}$ are their symmetric and
antisymmetric parts respectively. Although we are not with the riemannian connection
anymore, the parameter we have defined is such that $\theta$,
$\sigma_{\alpha\beta}$ and $\omega_{\alpha\beta}$ are still related to the expansion,
shear and rotation of the congruence respectively.\cite{Poulis-PhD}

\section{Variational Principle and Conformal Invariant Theory \label{sec - form varia e acopl}}
In order to make this geometric description compatible with a variational principle
for the dynamics of a particle in a gravitational field,\cite{Jurgen} we will restrict
ourselves to a particular case of Weyl geometry which is called Weyl Integrable
Space-time (WIST) in what follows. This is done simply by restricting the
field $\omega_\alpha$ to a gradient. That is, from now on we will consider only

\begin{equation}
\omega_\alpha \equiv \partial_\alpha \omega \, .
\label{omega em WIST}
\end{equation}

This kind of geometry is easily obtained by performing a Palatini approach to the action

\begin{equation}
S = \int e^{-\omega}R\sqrt{-g}d^4 x \, ,
\label{acao para Weyl - vacuo}
\end{equation}

\noindent where $R$ is the Ricci scalar and we follow the sign conventions of
Ref.~\refcite{Adler}. Variation of the connection gives precisely (\ref{conexao de Weyl})
with (\ref{omega em WIST}), what is necessary and suficient to give
(\ref{deriv da metrica em Weyl}) ensuring we are in WIST. Variation of the metric and
$\omega$ gives respectively:

\begin{eqnarray}
& & G_{\mu\nu} = 0 \, , \label{variacao da metrica - vacuo} \\
& & R = 0 \, , \label{variacao de omega - vacuo}
\end{eqnarray}

\noindent where $G_{\mu\nu}$ is the Einstein tensor. We see that variation of
$\omega$ gives a redundant equation, since (\ref{variacao de omega - vacuo}) is
the same as the trace of (\ref{variacao da metrica - vacuo}). Such equality implies
in a freedom to one of the functions to be determined. Soon it will be shown what this
freedom can imply in turn.

For test particles to obey the geodesic equation (\ref{Eq geodesica em Weyl}),
we first note that in WIST we have

\begin{equation}
u^\beta \omega_\beta = \frac{dx^\beta}{ds}\frac{\partial\omega}{\partial x^\beta} =
\frac{d\omega}{ds} \, ,
\label{prod de u com omega em WIST}
\end{equation}

\noindent so, the equation becomes

\begin{equation}
u^\beta\nabla_\beta u^\alpha =
\left( \frac{d}{ds}\ln\sqrt{u^\beta u_\beta} \right)u^\alpha -
\frac{1}{2}\frac{d\omega}{ds} u^\alpha =
 u^\alpha \frac{d}{ds}\ln\sqrt{e^{-\omega}g_{\mu\nu}u^\mu u^\nu} \, .
\label{Eq geodesica em WIST}
\end{equation}

\noindent Moreover, according to Refs.~\refcite{Chandra} and \refcite{Perlick},
in order to characterize a good clock for the observer who follows this geodesic,
 we should choose such a parameter in which the right hand side of the above equation
 is zero. This is achieved by precisely that same parameter $\tilde{s}$ defined before,
 in (\ref{parametro afim}). This equation is obtained by the following action for a
 test particle with mass $m$:

\begin{equation}
S_p = \int2m\int_{s_1}^{s_2}
\sqrt{e^{-\omega}g_{\mu\nu}\frac{dz^\mu}{ds}\frac{dz^\nu}{ds}}
\delta^4\left(x-z(s)\right)ds d^4x \, .
\label{acao para part teste em WIST}
\end{equation}

If we now notice that

\begin{eqnarray}
\Gamma^{\alpha}_{\beta\gamma} & = \frac{1}{2}\tilde{g}^{\alpha\lambda}
\left(\partial_\beta\tilde{g}_{\gamma\lambda} +
\partial_\gamma\tilde{g}_{\beta\lambda} -
\partial_\lambda\tilde{g}_{\beta\gamma} \right) \, ;
\label{conexao de Weyl como func de g til} \\
\tilde{g}_{\mu\nu} & \equiv e^{-\omega}g_{\mu\nu} \quad
\therefore \quad \tilde{g}^{\mu\nu} \equiv e^{\omega}g^{\mu\nu} \, ;\label{def de g til}
\end{eqnarray}

\noindent that is, the connection is just like the riemannian one
(Christoffel symbol) written with $\tilde{g}_{\mu\nu}$, we see that things work
much like they were in Riemann, but with $\tilde{g}_{\mu\nu}$ instead of $g_{\mu\nu}$
whenever it appears.

That is exactly what happened in the action (\ref{acao para part teste em WIST}) and
hence in the geodesic equation (\ref{Eq geodesica em WIST}). Specially if we take the
 parameter $s = \tilde{s}$ defined in Sec.~\ref{sec - congruencia de curvas}, which
 is again another case of replacing $g_{\mu\nu}$ by $\tilde{g}_{\mu\nu}$ in the
 definition of the affine parameter in the riemannian case. Moreover, the situation
 is the same in (\ref{acao para Weyl - vacuo}): since we have
 $\sqrt{-\tilde{g}} = e^{-2\omega}\sqrt{-g}$, the Einstein-Hilbert action becomes

\begin{align}
S & = \int R(\tilde{g}^{\mu\nu},\Gamma^\alpha_{\beta\gamma})
\sqrt{-\tilde{g}}d^4 x = \int \tilde{g}^{\mu\nu}R_{\mu\nu}(\Gamma^\alpha_{\beta\gamma})
\sqrt{-\tilde{g}}d^4 x = \nonumber \\
& = \int e^{\omega}g^{\mu\nu}R_{\mu\nu}(\Gamma^\alpha_{\beta\gamma})
e^{-2\omega}\sqrt{-g}d^4 x = \int e^{-\omega}R(g^{\mu\nu},
\Gamma^\alpha_{\beta\gamma})\sqrt{-g}d^4 x \, .
\label{acao para Weyl pela receita- vacuo}
\end{align}

Now we have made these considerations, it is quite clear that the most natural
choice for coupling other fields with $\omega$ is by keeping this same recipe.
That is, we take any other lagrangian of the form ${\cal L}\left(g^{\mu\nu}, ...\right)\sqrt{-g}$,
in GR, and write it as ${\cal L}\left(\tilde{g}^{\mu\nu}, ...\right)\sqrt{-\tilde{g}}$.
Then, we will have for our WIST action:

\begin{equation}
S = \int \left[R(\tilde{g}^{\mu\nu},\Gamma^\alpha_{\beta\gamma}) +
{\cal L}(\tilde{g}^{\mu\nu}, ...)\right]\sqrt{-\tilde{g}}d^4x \, .
\label{acao inv conf WIST}
\end{equation}

Variation of the connection gives us WIST. Variation with respect to the metric gives

\begin{align}
\frac{\delta S}{\delta g^{\mu\nu}} & =
\frac{\delta S}{\delta \tilde{g}^{\alpha\beta}}
\frac{\delta \tilde{g}^{\alpha\beta}}{\delta g^{\mu\nu}} =
\left[G_{\mu\nu}(\tilde{g}^{\mu\nu}) + T_{\mu\nu}(\tilde{g}^{\mu\nu}, ...)\right]
e^{\omega}\sqrt{-\tilde{g}} = 0 \quad \Rightarrow \nonumber \\
\Rightarrow \quad & G_{\mu\nu}(\tilde{g}^{\mu\nu}) =
 -T_{\mu\nu}(\tilde{g}^{\mu\nu}, ...) \;\; \therefore \;\; R(\tilde{g}^{\mu\nu}) =
  T(\tilde{g}^{\mu\nu}, ...) \, , \label{Eq Einstein inv conforme}
\end{align}

\noindent where we have defined

\begin{equation}
T_{\mu\nu}(\tilde{g}^{\mu\nu}, ...) \equiv
\frac{1}{\sqrt{-\tilde{g}}}\frac{\delta \left[{\cal L}\left(\tilde{g}^{\mu\nu}, ...\right)
\sqrt{-\tilde{g}}\right]}{\delta \tilde{g}^{\alpha\beta}} \, .
\label{tensor momento energia WIST}
\end{equation}

\noindent Varying $\omega$ we have

\begin{align}
\frac{\delta S}{\delta \omega} & =
\frac{\delta S}{\delta \tilde{g}^{\mu\nu}}\frac{\delta \tilde{g}^{\mu\nu}}{\delta \omega} =
 \left[G_{\mu\nu}(\tilde{g}^{\mu\nu}) + T_{\mu\nu}(\tilde{g}^{\mu\nu}, ...)\right]
 e^{\omega}g^{\mu\nu}\sqrt{-\tilde{g}} = 0 \quad \Rightarrow \nonumber \\
\Rightarrow \quad & R(\tilde{g}^{\mu\nu}) = T(\tilde{g}^{\mu\nu}, ...) \, ,
 \label{dinamica p omega WIST}
\end{align}

\noindent which is a redundant equation, since it is equal to (\ref{Eq Einstein inv conforme}).
So we still have a freedom to one of the functions to be determined. But now it is very
clear where this freedom comes from. Since the function $\omega$ always appears together
with the metric in the form $\tilde{g}_{\mu\nu}$, it is just a matter to see that it remains
unchanged under the transformation

\begin{equation}\left\{
\begin{array}{ccl}
g_{\mu\nu} & \rightarrow & \bar{g}_{\mu\nu} = e^{\Lambda}g_{\mu\nu} \, ; \\
\omega & \rightarrow & \bar{\omega} = \omega + \Lambda \, .
\end{array}
\right.
\label{transformacao metrica e omega}
\end{equation}

\noindent So we have an arbitrariness in the fields that keeps the theory just developed invariant.

Such invariance should be present in all observables defined on this theory, and so, our next task is to check it for the red-shift. We will now define it in this new context and show that it is invariant under the above transformation.

\section{Geometrical Optics and Red-Shift \label{sec - red-shift}}
Following the lines of Ref.~\refcite{Ellis} we have for the geometrical optics approximation
that light rays follow trajectories whose tangent vector, $k^\alpha$, is given by:

\begin{equation}
k^\alpha = g^{\alpha\beta}k_\beta \, , \quad k_\alpha \equiv \partial_\alpha\varphi \, ,
 \quad k^\alpha k_\alpha = 0 \, ,
\label{vetor de propagacao}
\end{equation}

\noindent where $\varphi$ is a phase. Moreover, according to
(\ref{contracao de X com deriv de X II}), we also have $k^\alpha\nabla_\beta k_\alpha = 0$.
Since $\nabla_\beta k_\alpha = \nabla_\alpha k_\beta$, we then have
$k^\alpha \nabla_\alpha k_\beta = 0$. This, in turn, implies

\begin{equation}
k^\alpha\nabla_\alpha\left(e^{\omega}k^\beta\right) = k^\alpha\hat{\nabla}_\alpha k^\beta = 0.
\label{vetor de prop geodetico em Riemann}
\end{equation}

\noindent That is, light rays follow null geodesics just like the manifold was riemannian
instead of a Weyl one.

For the red-shift we have the following relation between the frequency of the
electromagnetic waves in the moment of emission, $\nu_e$, and observation, $\nu_o$:

\begin{equation}
\frac{\nu_e}{\nu_o} = \frac{\left(k_\alpha u^\alpha\right)_e}{\left(k_\beta u^\beta\right)_o} \, .
\label{relac entre freq no redshift}
\end{equation}

The red-shift, $z$, of a source as measured by an observer is defined in terms of wavelenghts by

\begin{equation}
z \equiv \frac{\lambda_o - \lambda_e}{\lambda_e} \equiv \frac{\Delta \lambda}{\lambda_e}
\quad \Rightarrow \quad 1 + z = \frac{\lambda_o}{\lambda_e} = \frac{\nu_e}{\nu_o} =
 \frac{\left(k_\alpha u^\alpha\right)_e}{\left(k_\beta u^\beta\right)_o} \, ,
\label{def do redshift}
\end{equation}

\noindent which determines the red-shift from the vectors $u^\alpha_e$, $ u^\alpha_o$ and
the tangent vector $k^\alpha$ of the null geodesic. This relation is valid regardless of
the separation between the emitter and the observer and accounts for both Doppler and
gravitational red-shift.

Next, we make the following decomposition of $k^\alpha$: we consider an observer with
four-velocity $u^\alpha$ and let $n^\alpha$ be a projection of $k^\alpha$ into the
observer's rest frame, given by

\begin{eqnarray}
& & n^\alpha \equiv \frac{1}{u^\gamma k_\gamma}h^\alpha_\beta k^\beta \quad \Rightarrow
\quad n^\alpha n_\alpha = -e^{-\omega} \;\; , \;\; n^\alpha u_\alpha = 0 \, ,
\label{def de n ortogonal a u} \\
& & \therefore \quad k^\alpha = u_\beta k ^\beta\left(e^{-\omega}u^\alpha +
 n^\alpha\right) \, . \label{decomp de k em u e n}
\end{eqnarray}

Taking $v$ as the parameter along the null geodesic, we have $k^\alpha = dx^\alpha/dv$
and we can calculate the variation of $u^\alpha k_\alpha$ on an interval $dv$ along it as
being\footnote{For the case where both emitter and observer follow the unique fluid
velocity $u^\alpha$.}

\begin{equation}
d\left(u^\alpha k_\alpha\right) = \hat{\nabla}_\beta\left(u^\alpha k_\alpha\right)k^\beta dv
 = \left(\hat{\nabla}_\beta u_\alpha\right) k^\alpha k^\beta dv +
  u_\alpha \underbrace{k^\beta\hat{\nabla}_\beta k^\alpha}_{=0} dv \, .
\label{variacao infinitesimal de lambda 1}
\end{equation}

\noindent From (\ref{decomp irredutivel para V}) and (\ref{decomp de k em u e n})
we can rewrite it as

\begin{equation}
d\left(u^\alpha k_\alpha\right) = \left(\theta_{\alpha\beta}n^\alpha n^\beta +
 e^{-\omega}n^\alpha \dot{u}_\alpha \right)\left(u_\gamma k^\gamma\right)^2 dv \, .
\label{variacao infinitesimal de lambda 2}
\end{equation}

\noindent Now, from (\ref{def do redshift}) we have

\begin{equation}
\frac{d\lambda}{\lambda} = -\frac{d\left(u^\alpha k_\alpha\right)}{\left(u^\beta k_\beta\right)}
 = -\left(\theta_{\alpha\beta}n^\alpha n^\beta +
 e^{-\omega}n^\alpha \dot{u}_\alpha \right) u_\gamma k^\gamma dv \, ,
\label{variacao infinitesimal de lambda 3}
\end{equation}

\noindent which gives the variation of the wavelength along a small increment $dv$ on the
parameter that describes the light ray.

Next, we show that both this expression and (\ref{def do redshift}) are gauge invariants,
as expected.

\section{Verification of the Conformal Invariance of Red-Shift \label{sec - Verif inv conf}}

Let us examine the invariance of the objects that characterize the red-shift under
the conformal transformation (\ref{transformacao metrica e omega}). We see that the
following quantities from their definitions are transformed as follows:

\begin{align}
d\tilde{s} & = e^{-\omega}g_{\mu\nu}dx^\mu dx^\nu \rightarrow d\tilde{s} \, ;
\label{transformacao do parametro conforme} \\
u^\alpha & \equiv \frac{dx^\alpha}{d\tilde{s}} \rightarrow
u^\alpha \quad \therefore \quad u_\alpha = g_{\alpha\beta} u^\beta \rightarrow
 e^\Lambda u_\alpha \, ; \label{transformacao para u} \\
k_\alpha & \equiv \frac{\partial\varphi}{\partial x^\alpha} \rightarrow
k_\alpha \quad \therefore \quad k^\alpha = g^{\alpha\beta} k_\beta \rightarrow
 e^{-\Lambda} k^\alpha \, . \label{transformacao para k}
\end{align}

\noindent Thus, we have

\begin{equation}
u^{\alpha}k_{\alpha} \rightarrow u^{\alpha}k_{\alpha} \, ,
\quad u_{\alpha}k^{\alpha} \rightarrow u_{\alpha}k^{\alpha} \, ,
\label{transformacao de uk}
\end{equation}

\noindent and we confirm the invariance of (\ref{def do redshift}).
For equation (\ref{variacao infinitesimal de lambda 3}) we have

\begin{align}
h^\alpha_\beta & = \delta^\alpha_\beta - e^{-\omega}u^\alpha u_\beta \rightarrow
h^\alpha_\beta \quad \therefore \quad h_{\alpha\beta} \rightarrow
e^{\Lambda}h_{\alpha\beta} \;\; , \;\; h^{\alpha\beta} \rightarrow
 e^{-\Lambda}h^{\alpha\beta} \, ; \label{transformacao de h} \\
n^\alpha & = \frac{1}{\left(u^{\gamma}k_{\gamma}\right)}h^\alpha_\beta k^\beta
\rightarrow e^{-\Lambda}n^\alpha \quad \therefore \quad n_\alpha \rightarrow
 n_\alpha \, ; \label{transformacao de n}
\end{align}

\noindent and their different terms are transformed as:

\begin{align}
\theta_{\alpha\beta}n^\alpha n^\beta = n^\alpha n^\beta \nabla_\alpha u_\beta
 \quad & \rightarrow \quad e^{-2\Lambda}n^\alpha n^\beta
 \nabla_\alpha\left(e^\Lambda u_\beta\right) =
  e^{-\Lambda}\theta_{\alpha\beta}n^\alpha n^\beta \, ;
\label{transformacao da variacao infini de lambda 1} \\
\dot{u}_\alpha n^\alpha e^{-\omega} \quad & \rightarrow
 \quad e^{-\Lambda}\dot{u}_\alpha n^\alpha e^{-\omega} \, .
 \label{transformacao da variacao infini de lambda 2}
\end{align}

\noindent Given (\ref{transformacao de uk}), it remains only to determine the
transformation of $dv$, which is obtained from

\begin{equation}
k^\mu = \frac{dx^\mu}{dv} \rightarrow e^{-\Lambda}k^\mu \quad \Leftrightarrow
\quad dv \rightarrow e^\Lambda dv \, .
\label{transformacao de dv}
\end{equation}

\noindent So we have

\begin{multline}
\frac{d\lambda}{\lambda} = -\left(\theta_{\alpha\beta}n^\alpha n^\beta +
 e^{-\omega}n^\alpha \dot{u}_\alpha \right) u_\gamma k^\gamma dv \rightarrow \\
\rightarrow -\left(e^{-\Lambda}\theta_{\alpha\beta}n^\alpha n^\beta +
 e^{-\Lambda}e^{-\omega}n^\alpha \dot{u}_\alpha \right) u_\gamma k^\gamma
  e^\Lambda dv = \frac{d\lambda}{\lambda} \, ,
\label{transformacao da variacao infini de lambda 3}
\end{multline}

\noindent and we see that the gauge freedom of our theory still holds for the
 red-shift, preserving the arbitrariness to one of their geometrical objects.

\section{Consequences of Conformal Invariance in Cosmological Models
\label{sec - inv conf em mod cosmo}}

We have the Friedmann model for the three possible curvatures in a Weyl geometry
where, initially, $\omega = 0$:

\begin{equation}
ds^2 = d\tau^2 - a^2(\tau)\left[d\rho^2 + f^2(\rho)d\Omega^2\right]
\label{metrica de Friedmann em f(rho)}
\end{equation}

\noindent where $f(\rho)$ is equal to $\sin(\rho)$, $\rho$ or $\sinh(\rho)$
for the spherical, flat and hyperbolic cases, respectively. The red-shift can
be calculated from (\ref{relac entre freq no redshift}) or
(\ref{variacao infinitesimal de lambda 3}), where the velocities of the observer
and emitter are both equal to

\begin{equation}
u^\alpha = (1,0,0,0) \, .
\label{velocidade do observador em Friedmann}
\end{equation}

\noindent Now for $k^\alpha$, we have in the three models:

\begin{eqnarray}
& & k^\alpha = \frac{1}{a(\tau)}(1,\pm\frac{1}{a(\tau)},0,0) \quad \Leftrightarrow
\quad k_\alpha = \frac{1}{a(\tau)}(1,\mp a(\tau),0,0) \label{expressoes para k em Friedmann} \\
& & \therefore \quad k_\alpha u^\alpha = \frac{1}{a(\tau)} \quad \Rightarrow
\quad  1 + z = \frac{a(\tau_o)}{a(\tau_e)} \, . \label{prod de k com u em Friedmann}
\end{eqnarray}

This expression can also be obtained from (\ref{variacao infinitesimal de lambda 3}),
since we have in it:

\begin{equation}
\theta_{\alpha\beta}n^\alpha n^\beta = -\frac{1}{a(\tau)}\frac{da(\tau)}{d\tau} \, ,
\quad \dot{u}_\alpha = 0 \, , \quad u_\gamma k^\gamma dv = u_\gamma dx^\gamma =
d\tau \, , \label{red-shift em Friedmann 3}
\end{equation}

\noindent and the equation gets:

\begin{equation}
\frac{d\lambda}{\lambda} = \frac{1}{a(\tau)}\frac{da(\tau)}{d\tau} d\tau \;\;
\Rightarrow \;\; d\left(\ln\lambda\right) = d\left(\ln a\right) \quad \therefore \quad 1 + z = \frac{\lambda_o}{\lambda_e} = \frac{a(\tau_o)}{a(\tau_e)}
\label{z em Friedmann 2}
\end{equation}

However, being the metric (\ref{metrica de Friedmann em f(rho)}) conformally flat,
we can perform coordinate transformations in order to rewrite it as
$ds^2 = e^{-\Lambda}\eta_{\mu\nu}dx^\mu dx^\nu$ (see appendix). Then, we make the
transformation (\ref{transformacao metrica e omega}) and we are now with a Weyl
geometry where $g_{\mu\nu} = \eta_{\mu\nu}$, $\omega = \Lambda$ and we have the same
expression for the red-shift. That is, we now have the possibility that observations
of red-shift can be attributed to the field $\omega$ of a Weyl geometry in Minkowski space.
No longer indicating a curvature nor evolution to the metric. Likewise, we could have done
something in between, {\it i.e.}, a gauge transformation to get $\omega \neq 0$ and
$g_{\mu\nu} \neq \eta_{\mu\nu}$.

\section{Conclusions}
Motivated by the axiomatic approach of EPS and Woodhouse to characterize space-time
we performed a reformulation of GR in the context of WIST for the sake of consistency.

After introducing the variational formulation of their basic concepts, like the dynamics
for their geometrical objects and test particles, it was established how other physical
fields couples with the geometry. The particular coupling choosen endows our reformulation
with an unprecedented gauge symmetry that is of great theoretical interest by itself.
On the other hand, this same gauge freedom leaves the role of their geometric fields
undetermined. That is, one can choose any expression for one of them.

It was then presented a reformulation of the red-shift in which the symmetry is still
present. It was observed that in the cosmological case the gauge freedom may be used
to deprive the metric of any curvature or evolution. We can make, as a particular case,
such a transformation in which the metric becomes the Minkowski one and the whole red-shift
observations turns out to be described solely in terms of the new geometrical field, $\omega$.

We see that, so far, this reformulation still lacks a complete description of their
geometric objects. The attempt of having them fully described is our current subject
 of investigation.

\section*{Acknowledgments}
We acknowledge Conselho Nacional de Desenvolvimento Cient\'{\i}fico e Tecnol\'{o}gico for
financial support.

\appendix

\section{Coordinate Transformations for the Cosmological Case}
We want to perform coordinate transformations in order to rewrite the metric
(\ref{metrica de Friedmann em f(rho)}) as

\begin{equation}
ds^2 = e^{-\Lambda}\left[dt^2 - dr^2 - r^2 d\Omega^2\right] \, .
\end{equation}

\noindent Those are one for each case of the function $f(\rho)$, so we take
each of them separatelly.

\begin{itemize}

\item Flat case:

\begin{equation}
ds^2 = d\tau^2 - a^2(\tau)\left[d\rho^2 + \rho^2 d\Omega^2\right] =
e^{-\Lambda(t)}\left[dt^2 - dr^2 - r^2 d\Omega^2\right] \, .
\label{metrica de Friedmann caso plano}
\end{equation}

This is accomplished by making

\begin{eqnarray}
& & dt = \frac{d\tau}{a(\tau)} \, ; \label{caso plano - transf de coord t(tau)} \\
& & r = \rho \, ; \label{caso plano - transf de coord r(rho)} \\
& & e^{-\Lambda(t)} = A^2(t) \quad (A \equiv a\circ\tau) \, .
\label{caso plano - expressao para omega(t)}
\end{eqnarray}

\item Hiperbolic case

\begin{equation}
ds^2 = d\tau^2 - a^2(\tau)\left[d\rho^2 + \sinh^2(\rho) d\Omega^2\right] =
 e^{-\Lambda(\tau(t,r))}\left[dt^2 - dr^2 - r^2 d\Omega^2\right] \, .
\label{metrica de Friedmann caso hiperbolico-hiperbolico transf}
\end{equation}

\noindent This is achieved by the transformations:

\begin{equation}
\left\{\begin{array}{l}
r = e^{\frac{\Lambda(\tau)}{2}}a(\tau)\sinh(\rho) \, ; \\
t = e^{\frac{\Lambda(\tau)}{2}}a(\tau)\cosh(\rho) \, ;
\end{array}\right.
\quad \Leftrightarrow \quad
\left\{\begin{array}{l}
\rho = \tanh^{-1}\left(\frac{r}{t}\right) = \frac{1}{2}\ln\left(\frac{t+r}{t-r}\right) \, ; \\
e^{\frac{\Lambda(\tau)}{2}}a(\tau) = \sqrt{t^2 - r^2} \, .
\end{array}\right.
\label{caso hiperbolico - transf}
\end{equation}

\noindent With the condition that

\begin{equation}
\frac{d\Lambda(\tau)}{d\tau} = -\frac{2}{a(\tau)}\frac{d}{d\tau}\left[a(\tau)
\pm \tau\right] \quad \Leftrightarrow \quad \frac{d}{d\tau}
\left[e^{\frac{\Lambda(\tau)}{2}}a(\tau)\right] = \mp e^{\frac{\Lambda(\tau)}{2}} \, .
\label{caso hiperbolico - equacao para Lambda}
\end{equation}

\item Spherical case

\begin{equation}
ds^2 = d\tau^2 - a^2(\tau)\left[d\rho^2 + \sin^2(\rho) d\Omega^2\right] =
 e^{-\Lambda(\tau(t,r))}\left[dt^2 - dr^2 - r^2 d\Omega^2\right] \, .
\label{metrica de Friedmann caso hiperbolico-hiperbolico transf}
\end{equation}

This is done by simply making the changes:

\begin{equation}
\left\{\begin{array}{lcl}
a(\tau) \rightarrow \pm i a(\tau) \, ; \\
\rho \rightarrow \pm i \rho \, ;
\end{array}\right.
\label{caso esferico - transf}
\end{equation}

\noindent in the transformation used for the hyperbolic case, where $i = \sqrt{-1}$.

\end{itemize}

Given those transformations, it is straightforward to apply the gauge transformation
(\ref{transformacao metrica e omega}) and check that everything actualy tranforms like
it was shown in Sec.~\ref{sec - Verif inv conf}. The final result (\ref{z em Friedmann 2})
 then follows necessarily.

\end{document}